\documentclass[aps,pre,floatfix,showpacs]{revtex4}
\usepackage[]{graphicx}
\usepackage{amssymb,amsfonts,amsmath,longtable}

\begin{document}

\title{Coarsening in granular systems}

\author{Andrea Baldassarri}
\affiliation{Istituto dei Sistemi Complessi - Consiglio Nazionale delle Ricerche, Rome, Italy}
\affiliation{Dipartimento di Fisica, Universit\`a ``Sapienza'', Piazzale Aldo Moro 5, 00185 Rome, Italy}

\author{Andrea Puglisi}
\affiliation{Istituto dei Sistemi Complessi - Consiglio Nazionale delle Ricerche, Rome, Italy}
\affiliation{Dipartimento di Fisica, Universit\`a ``Sapienza'', Piazzale Aldo Moro 5, 00185 Rome, Italy}

\author{Alessandro Sarracino}
\affiliation{Sorbonne Universit\'es, UPMC Univ Paris 06, UMR 7600, LPTMC, F-75005, Paris, France}

\begin{abstract}
We review a few representative examples of granular experiments or
models where phase separation, accompanied by domain coarsening, is a
relevant phenomenon. We first elucidate the intrinsic non-equilibrium,
or athermal, nature of granular media. Thereafter, dilute systems, the
so-called ``granular gases'', are discussed: idealized kinetic models,
such as the gas of inelastic hard spheres in the cooling regime, are
the optimal playground to study the slow growth of correlated
structures, e.g. shear patterns, vortices and clusters.  
In fluidized
experiments, liquid-gas or solid-gas separations have been
observed. In the case of monolayers of particles, phase coexistence
and coarsening appear in several different setups, with
mechanical or electrostatic energy input. Phenomenological models
describe, even quantitatively, several experimental measures, both for
the coarsening dynamics and for the dynamic transition between
different granular phases.  The origin of the underlying bistability
is in general related to negative compressibility from granular
hydrodynamics computations, even if the understanding of the mechanism
is far from complete.  A relevant problem, with important industrial
applications, is related to the demixing or segregation of mixtures,
for instance in rotating tumblers or on horizontally vibrated
plates. Finally, the problem of compaction of highly dense granular
materials, which is relevant in many practical situations, is usually described
in terms of coarsening dynamics: there, bubbles of mis-aligned grains
evaporate, allowing the coalescence of optimally arranged islands and
a progressive reduction of total occupied volume.
\end{abstract}

\maketitle

\section{Introduction}

Granular systems are substances made of many {\em grains},
i.e. particles of average diameter roughly larger than $10^{-2}$
mm~\cite{andreotti}. The size of grains is such that interactions are
fairly modeled by dissipative hard core repulsion. Correspondingly,
granular systems are athermal, that is they do not posses neither a
spontaneous long term dynamics nor a thermodynamic equilibrium
state, except the trivial case of an inert immobile stack (or pile)~\cite{puglio,poeschel}.

However, injecting energy, usually by means of vibration, shaking,
tumbling or falling, leads the granular system to show a variety of
dynamical regimes, i.e. different ``phases'', roughly analogous to
solid, liquid or gas states of molecular matter~\cite{jaeger,K04}. It
is quite hard to push the analogy much forward, since dissipation, in
the form of tangential friction and inelastic collisions, makes
granular media intrinsically out of equilibrium: in many cases it is
evident that not only an Hamiltonian, but even a well-defined 
thermostat's ``temperature'' is lacking and therefore no Gibbs distribution can be postulated.

Notwithstanding the inherent non-equilibrium nature of granular
phases, many phenomena analogous to equilibrium phase transitions show
up in granular experiments and simulations. In most of them, a
variation of the input energy flux sensibly changes the internal
ordering of the material.  Sometimes, the transition from disordered
to ordered phase is associated with a growth in time of the size of
ordered domains, similarly to what happens in more standard coarsening
phenomena.  In this domain, an abrupt change of the energy input rate
plays the role of the usual quench in coarsening dynamics. In this
short review, we collect some noticeable examples where the concept of
coarsening is empirically meaningful for interpreting and
understanding results in the framework of granular systems. For a more
extensive review of patterns and collective behavior in granular
media, please refer to~\cite{AransonReview}.

The presentation follows a decreasing energy line. In
section~\ref{sec:cooling} we address the more dilute models, i.e. the
so-called granular gases, which in the cooling regime display
instabilities toward non-homogeneous states with growing
domains in the density and velocity field. Section~\ref{sec:mono}
concerns experiments with dilute or moderately dense shaken granular
materials, where several kinds of phase separation appear, with
ordered domains which slowly grow in time. The important case of
electrostatically driven granular fluids is discussed, with a few
noteworthy examples.  In section~\ref{sec:segr} we discuss the case of
demixing or segregation, which usually applies to dense granular
materials, slowly agitated or rotated in drums. In
section~\ref{sec:comp}, the compaction dynamics, which is often
interpreted as an evaporation of alignment defects or a coarsening of
aligned domains, is briefly reviewed. Finally, the last section draws
conclusions and perspectives.

\section{Cooling granular gases}
\label{sec:cooling}
Fluidization of granular media is achieved by injecting mechanical
energy into the system, typically by shaking the whole container or
vibrating one of its sides~\cite{puglio}. When packing fraction $\phi$
is low enough (typically lower than $50\%$), a gas-like or liquid-like
stationary state is rapidly achieved, characterized by a ``granular
temperature'' which is defined as $T_g=\frac{1}{d}m\langle v^2
\rangle$ where $d$ is the dimensionality, $m$ the mass of a grain and
$v^2$ the squared modulus of its vectorial velocity. The granular
temperature is given by a balance between the energy injected and the
dissipation in collisions, which is usually parametrized by a
restitution coefficient $\alpha \le 1$ ($\alpha=1$ for elastic
collisions). Several examples of ``phase-transitions'' have been
recognized in fluidized granular systems. In the absence of an
interaction energy scale, due to the hard-core nature of the
grain-grain collisions, the transition is usually controlled by
packing fraction, or by restitution coefficient, rather than the
granular temperature. A relevant exception is constituted by the
sudden quench protocol, where the fluidizing mechanism is abruptly
interrupted and a ``cooling'' regime intervenes. In this cooling
regime, typically studied in simulations and within kinetic
theory~\cite{poeschel}, the growth of ordered structures in the
velocity field (vortices or shear bands) and in the density field
(clustering) is observed~\cite{noije}. In kinetic theory the idealized
starting point is the so-called Homogeneous Cooling State (HCS), which
is a spatially homogeneous solution of the inelastic Boltzmann
equation where the temperature follows the Haff's law,
i.e. asymptotically $T_g(t) \sim t^{-2}$. Granular Hydrodynamics
(GH)~\cite{brey}, which is expected to describe the evolution of
``slow fields'' (density, macroscopic velocity and granular
temperature), is the simplest theory to predict the instability of the
HCS: near the HCS one may linearize the system of GH equations and, in
space-Fourier transform, find a linear algebraic system for each
wave-vector $k$. Eigenvalues of the system are guaranteed to be
negative only for large enough $k$: for $k<k_\perp$ the shear mode
(rescaled by $\sqrt{T_g}$) becomes unstable and structures such as
vortices display a correlation length growing as $\sim \sqrt{\tau}$
(where $\tau$ is the time measured in cumulated number of collisions),
while for $k<k_\parallel$ a mode involving density becomes also
unstable, growing with a similar law (cluster
formation)~\cite{noije}. Since $k_\perp \sim \sqrt{1-\alpha^2}$ while
$k_\parallel\sim 1-\alpha^2$, one has $k_\perp \ll k_\parallel$ for
$\alpha \to 1$. It is possible, for instance, to choose a linear size
$L$ of the system such that
$k_\perp>k_{min}=\frac{2\pi}{L}>k_\parallel$, which implies that the
shear mode is unstable but the density mode is not: in the linear
(initial) stage there is only the appearance of shear structures in
the velocity field, while no clustering is observed.

\begin{figure}
\includegraphics[width=6cm,angle=-90]{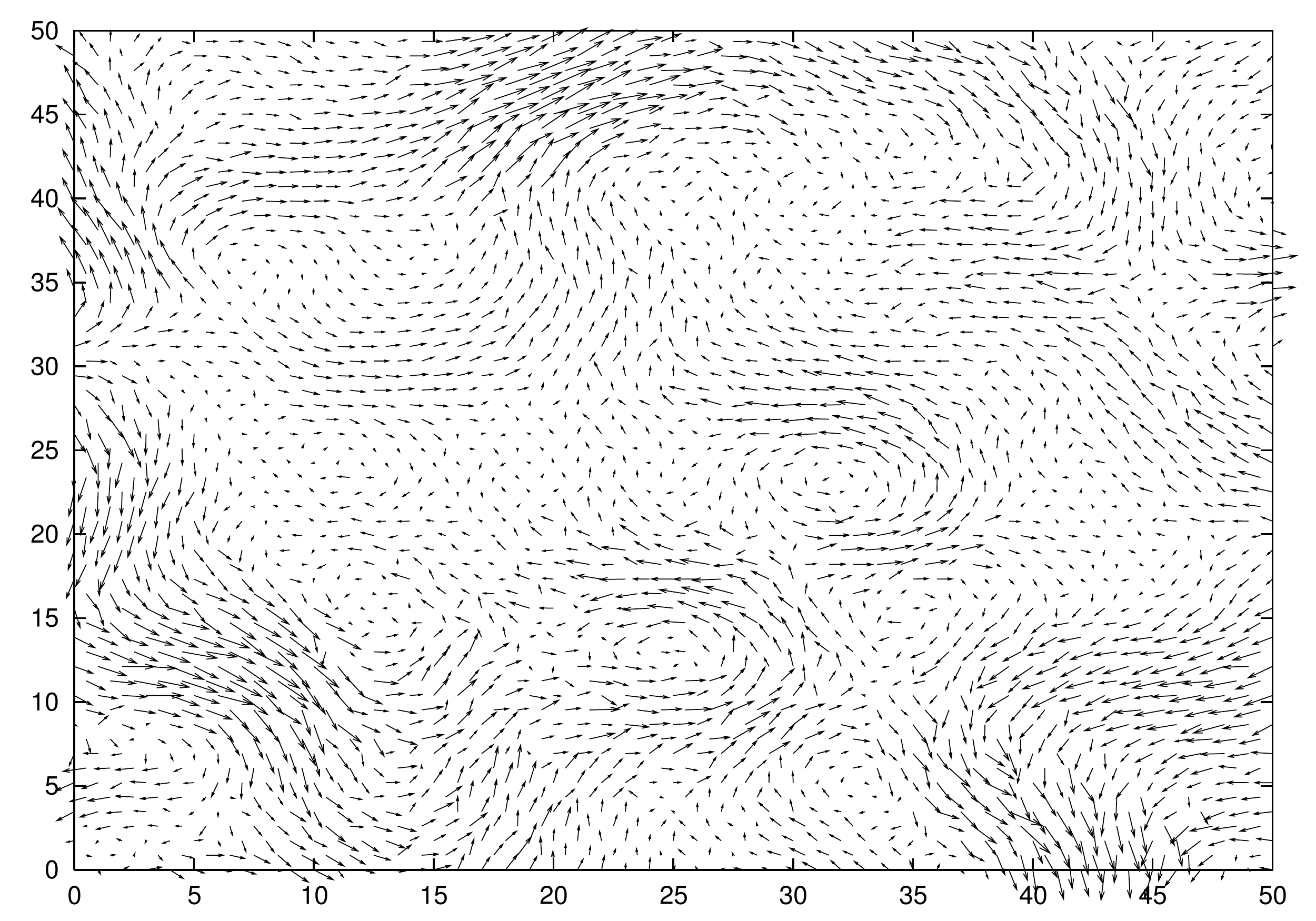}
\includegraphics[width=6cm,angle=-90]{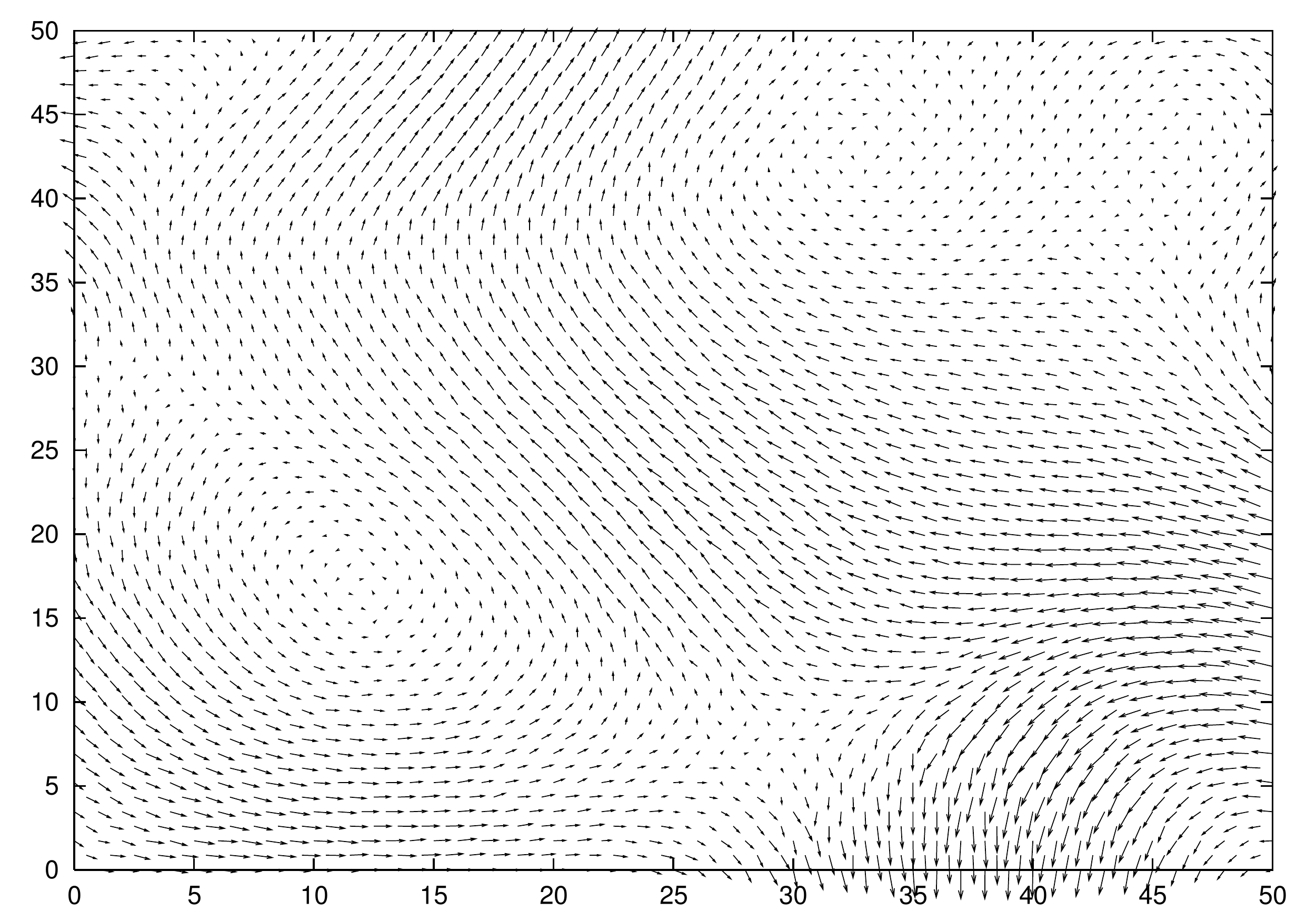}
\caption{Coarsening of vortices in the 2d version of the lattice model for the incompressible cooling state, with size of the system $512 \times 512$ and restitution coefficient $\alpha=0.7$~\cite{bald1}. The two plots represent two different times, measured in number of collisions per particle: Left: $\tau=52$, Right: $\tau=535$.\label{fig:vortici}}
\end{figure}

After some time from the onset of instabilities, the system enters a
non-linear regime which is no more described by linear GH. The study
of full GH equations is difficult and gives place to many possible
regimes depending on geometry, dimensionality, boundary conditions,
degree of inelasticity, etc~\cite{meerson}.  Some observations with Molecular
Dynamics (MD) in $1d$ showed that the cooling system asymptotically
reaches a regime where energy decays as $t^{-2/3}$ which is
independent of $\alpha$ and is therefore equivalent to the dynamics of
the sticky gas ($\alpha=0$)~\cite{ben-naim}. The sticky gas, characterized by a
velocity field with traveling shocks and a density field with
coalescing clusters, is well described by the inviscid Burgers
equation and is consistent with the observed energy decay. It appears,
however, that this scenario breaks down at $d>1$, since there the $\alpha=0$
case is no more equivalent to the sticky gas. A series of
idealized models on the lattice have been proposed to study the
``incompressible'' dynamics of the velocity field, i.e. assuming that the
density does not change appreciably from the initial homogeneous
configuration~\cite{bald2,bald1}. On each site of the lattice (1d or 2d squared lattice)
there is a particle which can collide with its neighbors, dissipating
energy and conserving total momentum. For simplicity the dependence of
the collision rate upon the relative velocity has been neglected,
while the so-called kinematic constraint, which forbid collisions
between particles going in opposite directions in their
center-of-mass reference frame, has been retained. Comparison with MD
results in 1d is interestingly good~\cite{bald2}, even for the clustering regime,
by {\em unrolling} the coordinates i.e. replacing real position with
particle's index, which is analogous to a sort of Lagrangian
coordinate. The study of velocity correlations, through the structure
factors, reveals a correlation length $\xi$ which grows as $\sim
1/T_g(t)$. If time is measured by the cumulated number of collisions
$\tau$, one discovers that $T_g \sim \tau^{-1/2}$ and $\xi \sim
\tau^{1/2}$, as in a simple diffusive process. In $2d$ the lattice
model remains consistent with the previous diffusive scenario~\cite{bald1},
revealing an energy decaying as $\tau^{-d/2}=\tau^{-1}$ and a correlation
length growing again as $\tau^{1/2}$. This correlation length in $2d$
is clearly associated with the growth of vortices (see Fig.~\ref{fig:vortici}).

\section{Phase coexistence in driven monolayers}
\label{sec:mono}

The study of granular cooling can be considered as a theoretical
framework useful to set up new tools and concepts for non-equilibrium
physics. Experimentally, it is much more relevant to study the
stationary dynamics resulting from a moderate agitation of the grains.
In such a case, correlations usually increase with the density of the
system. The dynamics of fluidized granular beds has shown a rich
variety of interesting ordering phenomena, mainly the appearance of
surface patterns, localized waves and excitations
(e.g. ``oscillons'').  For a small number of particles, the system
reduces to an effective 2D geometry, and there is no more distinction
between surface and bulk: this is the optimal setup where a
phase-separation scenario clearly emerges.

In a seminal paper, Olafsen and Urbach~\cite{Olafsen1998} performed
experiments with inelastic particles (stainless steel balls) on a
horizontal aluminum plane. When the number of particles is
insufficient to complete a single layer, the system is named a {\em
  submonolayer} of particles.  Such a system can be gently shaken
imposing a sinusoidal vertical displacement of the plate
($z(t)=A\sin(\omega t)$).  The relevant parameter is the dimensionless
acceleration $\Gamma=A\omega^2/g$, where $g$ is the gravitational acceleration.

\begin{figure}
\includegraphics[width=8cm]{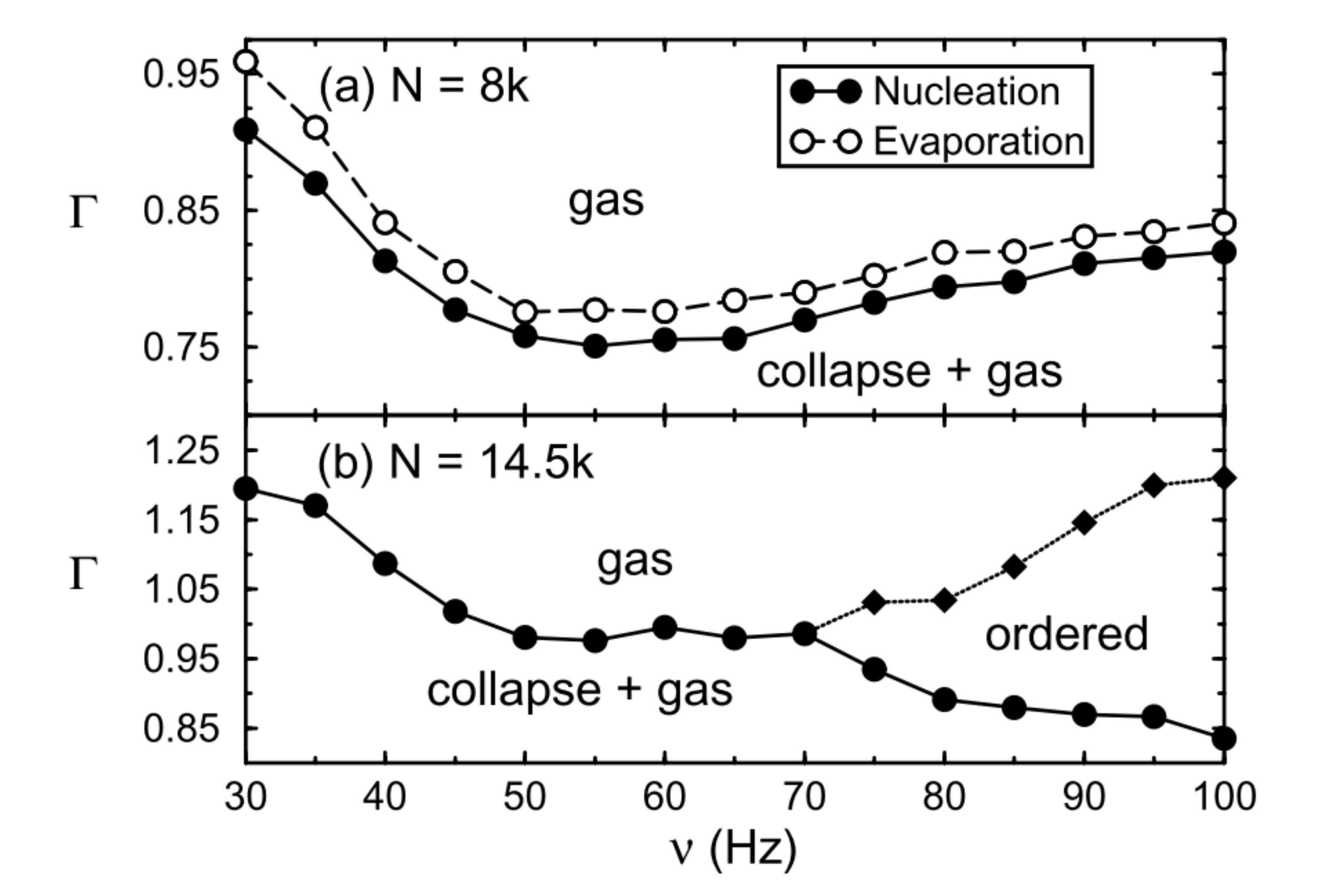}
\caption{The phase diagrams for the vibrated monolayer experiment
  in~\cite{Olafsen1998}. Two sets of experiments with different number
  of particles are considered: (a) $N = 8000$ particles and (b)
  $N=14500$ particles. The filled circles denote the acceleration
  where the collapse nucleates. The open circles in (a) indicate the
  point where the collapse disappears upon increasing the
  acceleration. The diamonds in (b) show the transition to the ordered
  (hexagonal) state as the acceleration is reduced. Reprinted figure
  with permission from~\cite{Olafsen1998}. \label{fig:olafsen}}
\end{figure}

The regime of interest is when $\Gamma$ is not too large, so that
particles cannot hop over one another and the motion is effectively
on a two dimensional layer. Within this regime, the system
behaviour depends on $\Gamma$, as well as on the vibration angular frequency
$\omega$ (or the vibration frequency $\nu=\omega/2\pi$) and on the total number
of particles $N$ in the system (see Fig.~\ref{fig:olafsen}).

For very low $\Gamma$, irrespectively of the frequency $\nu$, a condensate of particles at rest on the plate, while
in contact with each other, appears. For small number of particles, such structure nucleates as an island  surrounded by
rapidly moving particles (see Fig.~\ref{fig:olafsen}, top panel). In systems with a large number of particles
(but still submonolayer), ordered clusters of moving particles may
also appear, when shaken at high frequency (see Fig.~\ref{fig:olafsen}, bottom panel). Particles in such clusters
move around an average position, disposed in a regular (hexagonal)
lattice. The study of the quite
interesting phase diagram manifests several analogies with
liquid-solid transition, with phase coexistence and histeretic
features. Several other
experiments~\cite{Losert1999,Prevost2004,Olafsen2005} confirmed the
phase coexistence scenario, suggesting the possibility of phase
ordering kinetics and coarsening of clusters in vibrofluidized
granular submonolayers. Recently, pattern formation in submonolayers
{\em horizontally shaken} has also been observed and
investigated~\cite{Krengel2013}. The formation of strike-like patterns
in monodisperse submonolayers, reproduced in molecular dynamics
simulation, is quite fast (about $10$s) and no real coarsening
dynamics can been appreciated, at odds with the case of binary
mixtures (see below).

In order to better investigate similar phenomena, a different set up
has been proposed~\cite{Aranson2000} for much tiny (about $40\mu m$
diameter) metallic particles contained between two horizontal metallic
plates. The energy was electrostatically injected applying an
oscillating voltage between the plates. Quite independently from the
oscillating frequency, two threshold values for the amplitude of the
resulting electric field appear.  Above a first, lower value, $E_1$,
isolated particles detach from the bottom plate and start to
bounce. Above an higher value, $E_2$, all the particles move and the
granular medium forms a uniform gas-like phase.

\begin{figure}
\includegraphics[width=11cm]{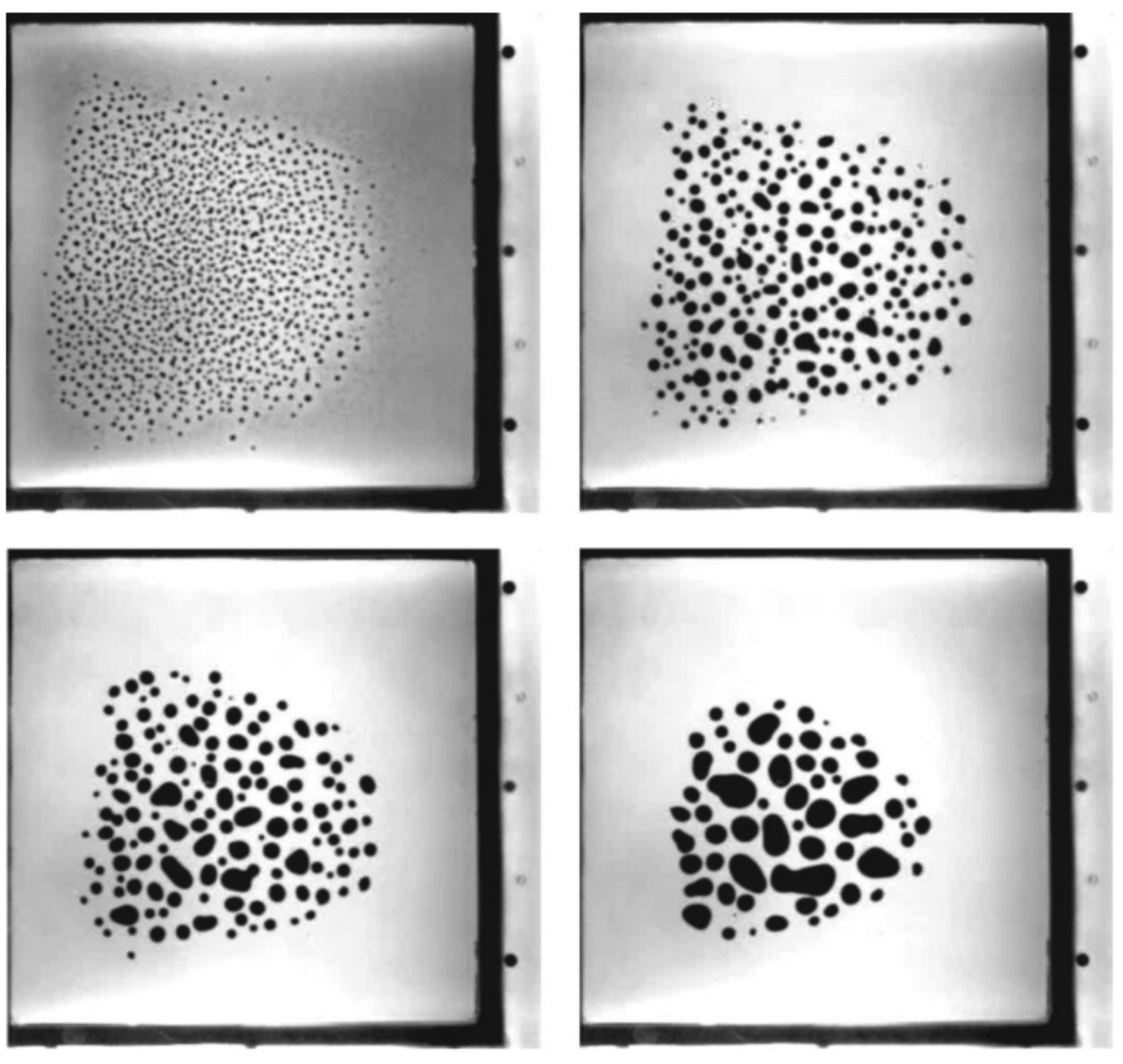}
\caption{Snapshots of coarsening of granular clusters for experiments with tiny particles (about $40\mu m$
diameter) in an electro-static cell at times $t=0s$ (upper left), $t=10^4s$ (upper
  right),$t=2\cdot 10^4s$ (lower left), and $t=5\cdot 10^4s$ (lower
  right). The applied dc electric field $E=2.33$ kV/cm. 
   Reprinted figure with permission from~\cite{Sapozhnikov2005}. A movie is avalaible as supplementary material of the original paper~\cite{Sapozhnikov2005}. Copyright (2005) by the
American Physical Society.\label{ab:electrocoarsening}}
\end{figure}

Interestingly, decreasing the field from such a highly mobilized phase
to a value comprised between the two thresholds $E_1<E<E_2$, a
phenomenon analogous to coalescence dynamics is observed (see
Fig.~\ref{ab:electrocoarsening}). The phenomenon is analogous to the
one observed in vertically vibrating steel spheres described above. In
fact, a set-up where the same granular medium could be either excited
mechanically or electrostatically has been
investigated~\cite{Sapozhnikov2003}. The appearance of the same
ordered "solid-like" phase (clusters of immobile particles) is
reported when the energy injection rate is decreased, both for the
vibrated and for the electrostatically driven case.

The extremely small size of the grains allows one to study the
dynamics of a very large number of particles. Just after the quench,
many "solid" clusters start to form. Their number decreases in time as
$N(t)\propto 1/t$. On the other hand, their average surface grows as
$\langle \Sigma(t)\rangle\propto t$, as long as the dynamics keeps quasi
2D. (For low frequency of the applied field a different large time
growth $\langle \Sigma(t)\rangle \propto t^{2/3}$ is observed. In this
regime the grains can easily hop on each other, the clusters contain
more than a monolayer of the particles, and the motion is effectively
3D).

A phenomenological model has been proposed
in~\cite{Aranson2000,Aranson2002} to describe these experiments. The
local density $n$ of the precipitate phase (the density of immobile
particles) is supposed to evolve according to a phenomenological
equation
\begin{equation}
\partial_t n = \nabla^2 n +\phi(n,n_g),
\label{ab:aransonmodel}
\end{equation}
where $n_g$ is the number density of bouncing particles
(i.e. particles of the gas phase). 
The function $\phi$, at fixed $n_g$, has two stable zeros as a function of $n$, corresponding to
$n=0$ (gas) and $n=1$ (solid), separated by an unstable zero in
between. The function $\phi$ in Eq.~(\ref{ab:aransonmodel}) characterizes the
solid-gas conversion rate. The effectiveness of solid-gas transitions
is controlled by the local gas concentration $n_g$. 
In other words, at fixed $n_g$, Eq.~(\ref{ab:aransonmodel}) is similar to a
time-dependent Ginzburg Landau Equation (GLE), whose solution, from an
initial random configuration, displays a coarsening dynamics of the
solid domains, where $n\simeq 1$. The driving mechanism of the
coarsening is the progressive reduction of the curvature of the domain
interfaces, as described by the Allen-Cahn equation~\cite{B94}. This
scenario is named ``model A'' in the classification of Hohenberg and
Halperin~\cite{Hohenberg77}. In this case the average domain size
grows as $\langle R(t)\rangle \propto \sqrt{t}$ and their number decreases as
$N(t)\propto 1/t$.

However, here $n_g$ is not necessarily constant, neither in space, nor in
time, and we should provide an equation for its time evolution to
couple with Eq.~(\ref{ab:aransonmodel}).  Nevertheless, exploiting the observation
that density relaxation in the gas phase is fast compared to the
cluster-gas exchange dynamics, one can assume that $n_g$ is
approximately constant in space and depends only on time.  On the
other hand, global density of the system is constant in time, hence
\begin{equation}
\int \, (n+n_g)\,dx\,dy = S n_g(t) +  \int \,n\,dx\,dy = M,\label{ab:conservation}
\end{equation}
where $S$ is the surface of the plate and $M$ the total number of particles.
It turns out that $n_g(t)$ plays the role of a ``mean field''
interaction between gas and solid, and, on long time scales, it
becomes close to a special ``equilibrium'' value $n^{eq}_g$. 
Hence, due to the presence of $n_g$, the phenomenological
model~\cite{Aranson2002} does not correspond to the model A. At odds
with model A, here the coarsening field has to satisfy the
conservation constraint given by Eq.~(\ref{ab:conservation}).

In general, conservation properties of the coarsening field affects
the dynamics.  When the field is conserved, larger domains can grow
only at the expenses of smaller ones. For instance, if the coarsening
field is locally conserved (which is the case of a dynamics described
by a Cahn-Hillard equation) this can happen through transfer of solid
between interfaces via bulk diffusion~\cite{Lifshitz61}. The scaling
exponents of the average domain size in this case are different (model
B in the classification of Hohenberg and Halperin~\cite{Hohenberg77}):
$\langle R(t)\rangle\propto t^{1/3}$.

Nevertheless, the conservation given by Eq.~(\ref{ab:conservation}) does
not have such a dramatic effect.  For long time, when $n_g$ attains a
constant equilibrium value, this condition resembles that of a
globally conserved order parameter: $\int \,n(x,y)\,dx dy =
\mbox{constant}$. In this case, the scaling behavior predicted by
model A is recovered: $\langle R(t)\rangle \propto \sqrt{t}$.
According to the mean field theory Lifshitz-Slyzof-Wagner
(LSW)~\cite{Lifshitz61,Wagner61}, the reason for the recovering of
Allen-Cahn scaling, is that the normal velocity of interfaces are
given by the excess curvature with respect to the interface curvature
averaged over the whole interface in the system (this mechanism is
named Ostwald ripening for interface controlled
dynamics~\cite{Wagner61}).

In order to corroborate the phenomenological model~\cite{Aranson2002}
and its connection with the coarsening with global conservation, the
size distribution of clusters has been considered. Experimental
investigation in the coarsening dynamics of submonolayer granular
media~\cite{Sapozhnikov2005}, allowed the experimental measure of the
(scaling) size distribution of clusters. Apparently, the observed
distributions, however, do not agree with the theoretical prediction
of the mean field LSW theory~\cite{Wagner61}. In fact, Conti et
al.~\cite{Conti2002} showed that coalescence of neighboring clusters
should not be neglected, as the mean field treatment did.  Taking into
account such a correction, a good agreement is obtained both with
numerical simulation of the global conserved GLE model, and with the
observation in the granular experiment~\cite{Sapozhnikov2005}.

\begin{figure}
\includegraphics[width=11cm]{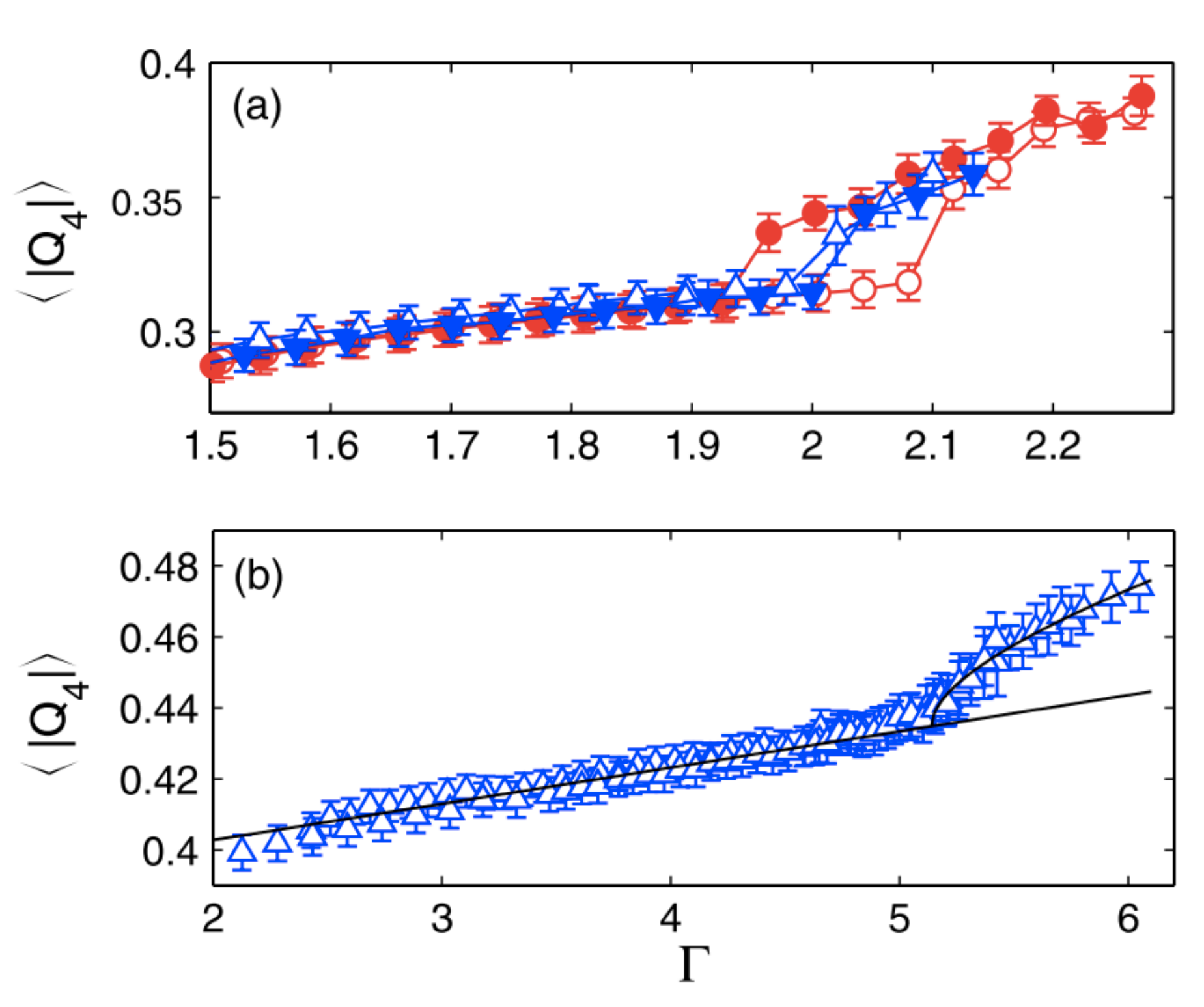}
\caption{ Average global fourfold bond-orientational order parameter
  $Q_4$ versus relative acceleration $\Gamma$. Upper panel (a) shows
  the case of first-order phase transition (with hysteresis
  effects). Lower panel (b) shows a continuous transition.  Continuous
  lines in (b) correspond to fits a linear trend for the sub-critical
  region $2.5<\Gamma<5$, to which has been over-imposed a power law in
  the supercritical region $\Gamma > 5$ (with a resulting exponent of
  $1/2$).  Reprinted figure with permission from~\cite{Castillo2012}. Copyright (2012) by the
American Physical Society.\label{ab:q4}}
\end{figure}

Recent interesting developments are reported in two papers by Castillo
et al.~\cite{Castillo2012,Castillo2015}. Experiments are performed in
vertically vibrated monolayers of $1$mm size steel particles, close to
the "solid-liquid" transition. The measure of the system structure
factor shows that density fluctuations increase in size and intensity
as the transition is approached, but they do not change significantly
at the transition itself. On the other hand, the dense, metastable
clusters, increase their local order in the vicinity of the
transition. Exploiting the square symmetry appearing in the dense
phase for the specific set-up in study, a bond-orientational order
parameter ($Q_4$) has been defined. This quantity evidences a critical
behavior at the transition, behaving as a first- or second-order
phase transition, depending on filling density and vertical
height~(see Fig.\ref{ab:q4}). In the case of continuous transition,
power law has been observed for correlation length, relaxation time
and static susceptibility of the order parameter $Q_4$. The corresponding
critical exponents are consistent with model C in the Hohenberg and
Halperin classification~\cite{Hohenberg77} of dynamical critical
phenomena, valid for a non-conserved critical order parameter
(bond-orientation order) coupled to a conserved field (density).

Should these results be confirmed, the observed dynamical phase
transition would determine the corresponding coarsening dynamics
after a quench.  Interestingly, model C displays a quite complex
coarsening scenario, with different growing laws for different
quenching procedures~\cite{Kockelkoren2002}.

The origin of bistability and phase separation in vibrated granular
monolayers, in terms of an hydrodynamic description, is not yet fully
understood. Several numerical simulations and experiments have been
designed in order to advance in this direction.

\begin{figure}
\includegraphics[width=11cm]{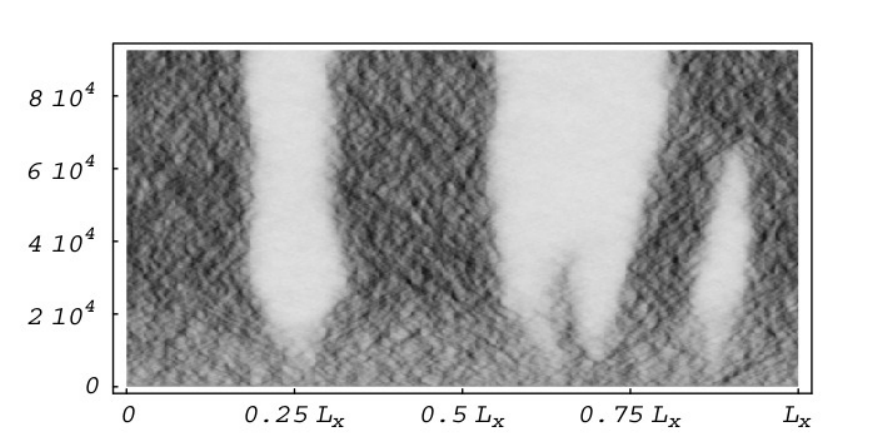}
\caption{Spatiotemporal evolution of the y-averaged density field,
  with time on the vertical axis and increasing upward. The gray scale
  is proportional to density, with darker regions representing denser
  regions in the system. Reprinted figure with permission from~\cite{soto}. Copyright (2002) by the
American Physical Society.\label{vdw}}
\end{figure}

A transition between a gas-like and liquid-like state with bubble
nucleation and subsequent coarsening has been observed in simulations
in a vibrated box~\cite{soto}, see Fig.~\ref{vdw}. This transition has been explained by
means of a van der Waals-like macroscopic theory: in such a theory the
phase coexistence is usually guaranteed by a nonmonotonous behavior of
the pressure as a function of the density, which implies the presence
of a bistable region separated by a coexistence and a spinodal
curve. The inflection point of pressure versus density (negative
compressibility), for granular particles, is due to the energy balance
between energy injection (shaking) and energy dissipated in
inter-particle collisions, which leads to a stationary $T_g$, 
decreasing with density.

A partial confirmation of this scenario has emerged in a quasi 1d
vibrated experiment~\cite{clerc}. In the experiment phase separation
of clusters and coalescence through coarsening has been obtained.  The
clusters are in a crystal-like (square or triangular lattice) phase
and coexist with a fluid phase. Even if the nature of the two phases
is different, a similar van-der-Waals like theory has been used to
explain the experimental result. Measurement of pressure have
indicated the existence of a pressure plateau corresponding to the
coexistence region. Slow coarsening and coalescence of nucleated
crystal islands is observed in time. 

Starting from Navier-Stokes granular hydrodynamic equation, Khain et
al.~\cite{Khain2011} proposed a different scenario, where the
bistability results from the nontrivial energy injection mechanism
from a vibrating plate to the granular gas. This is different from the
mechanism advocated in~\cite{soto}, where the van-der-Walls instability
is rather attributed to energy loss due to inelastic collision between
particles. In both cases, the phase separation is associated with a
negative compressibility of the granular gas.

\begin{figure}
\includegraphics[width=11cm]{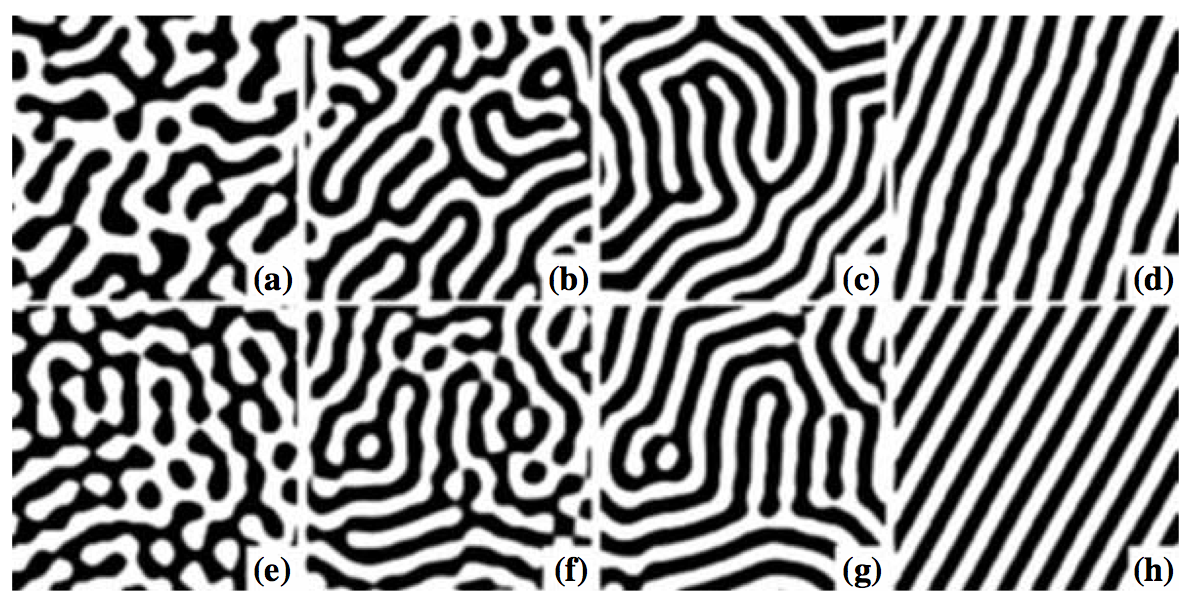}
\caption{Spatiotemporal evolution of the patterns during experiments
  and simulations described in~\cite{KP02}.  Images (a)-(d): The real
  images of the free surface of granular layers after a sudden quench
  from $\Gamma=2.4$ to $\Gamma= 2.8$: times (in period of vibration)
  are $t=2$ (a), $t=10$ (b), $t=200$ (c), and $t=1000$ (d). The bright
  parts correspond to the crests of the free surface, and the dark
  parts correspond to the troughs of the free surface.  (e)-(h): The
  images of the numerical results of the 2D Swift-Hohenberg equation
  in time (see~\cite{KP02} for parameters of the simulation).
  Reprinted figure with permission from~\cite{KP02}. Copyright (2002)
  by the American Physical Society.\label{fig:kp}}
\end{figure}

Another example of coarsening in vibrated granular media, where a
mono-disperse granular material is studied, is shown in the experiment
described in~\cite{KP02}. Here the formation and evolution of
regular patterns in a vertically vibrated thin granular layer of
phosphor-bronze spherical beads is analyzed.
This system shows a rich phenomenology: an initially flat layer can
form complex structures, such as stripes, squares, hexagons, etc, see for instance Fig.~\ref{fig:kp}. The
analysis focuses of the coarsening dynamics following a sudden change
of a control parameter (the vibration amplitude), which leads to
stripe structure. The first stage of this evolution is characterized
by the formation of rolls of well-defined wavelength with random
orientation. Later, these stripes are observed to align and an ordered
structure is eventually attained. This behavior can be analyzed within
the Swift-Hohenberg theory~\cite{SH77}, which describes the dynamics
of the order parameter in nonlinear dissipative systems, with
instability of the first kind (namely, the instability grows only at a
finite wave vector.) This model predicts a power-law growth for the
characteristic length $L(t)\sim t^z$ with $z=1/4$. In the granular
experiment the order parameter is defined as a function of the local
stripe orientations $\theta(\boldsymbol{r},t)$, which is defined
  as the local angles of the stripes ($\boldsymbol{r}$ denoting the
  position). This allows one to define the local order parameter
  $\psi(\boldsymbol{r},t)=\exp[2i\theta(\boldsymbol{r},t)]$ and to
  measure the angle averaged correlation function $C(r,t)=\langle
  \textrm{Re}[\psi(0)^*\psi(r)]\rangle$.  This satisfies a scaling
  relation with the dynamic correlation length in the system
  $C(r,t)=g[r/L(t)]$, where $L(t)\sim t^z$ is the characteristic
  length scale of the ordered domain at time $t$. The time behavior
of this length turns out to be in agreement with the Swift-Hohenberg
model.

These examples show that tools and principles of standard statistical
mechanics, developed in the context of phase transition and ordering
kinetics, can be effective, in some cases, for the description of
athermal systems out of equilibrium.

\section{Segregation of mixtures via coarsening}
\label{sec:segr}
A relevant problem for dense granular systems, with important
industrial applications, is the control of mixtures of different
grains, which can differ for sizes, densities, shapes, friction
coefficients, etc. A large amount of experimental and numerical work
has been devoted to the study of demixing or segregation. This is a
huge field and we refer the reader to the review article~\cite{K04}
for experiments and theories on spontaneous segregation in granular
mixtures.  Here we focus on the cases where such a phenomenon occurs
via domain coarsening of the components and present results for two
classes of experiments: vibrated two-dimensional granular systems and
rotating tumblers.

\paragraph{Quasi two-dimensional layers.} The simplest setup to study the phenomenon of spontaneous segregation
is a quasi two-dimensional configuration, where the granular mixture
is spread above a horizontally vibrated plate. In this case, several
studies have shown that mixture segregation occurs through stripe
coarsening, which develop orthogonally to the vibration direction, and
is due to the differential frictional drag acting on the different
components of the mixture.

\begin{figure}
\includegraphics[width=3cm,clip=true]{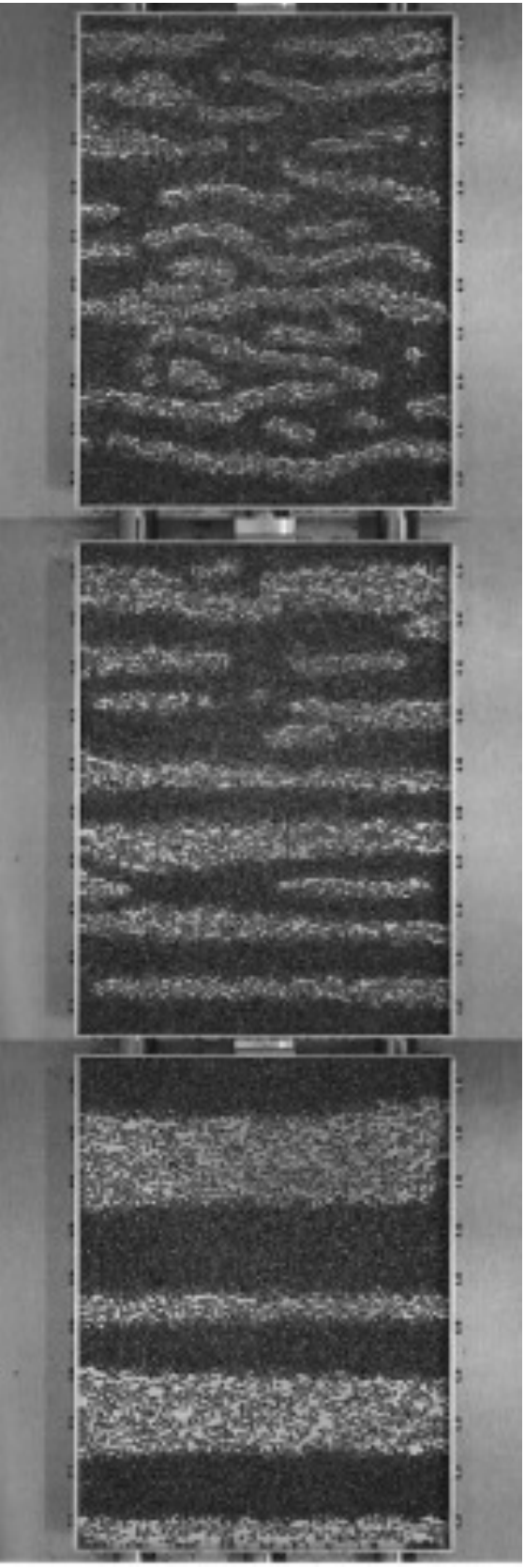}
\includegraphics[width=5cm,clip=true]{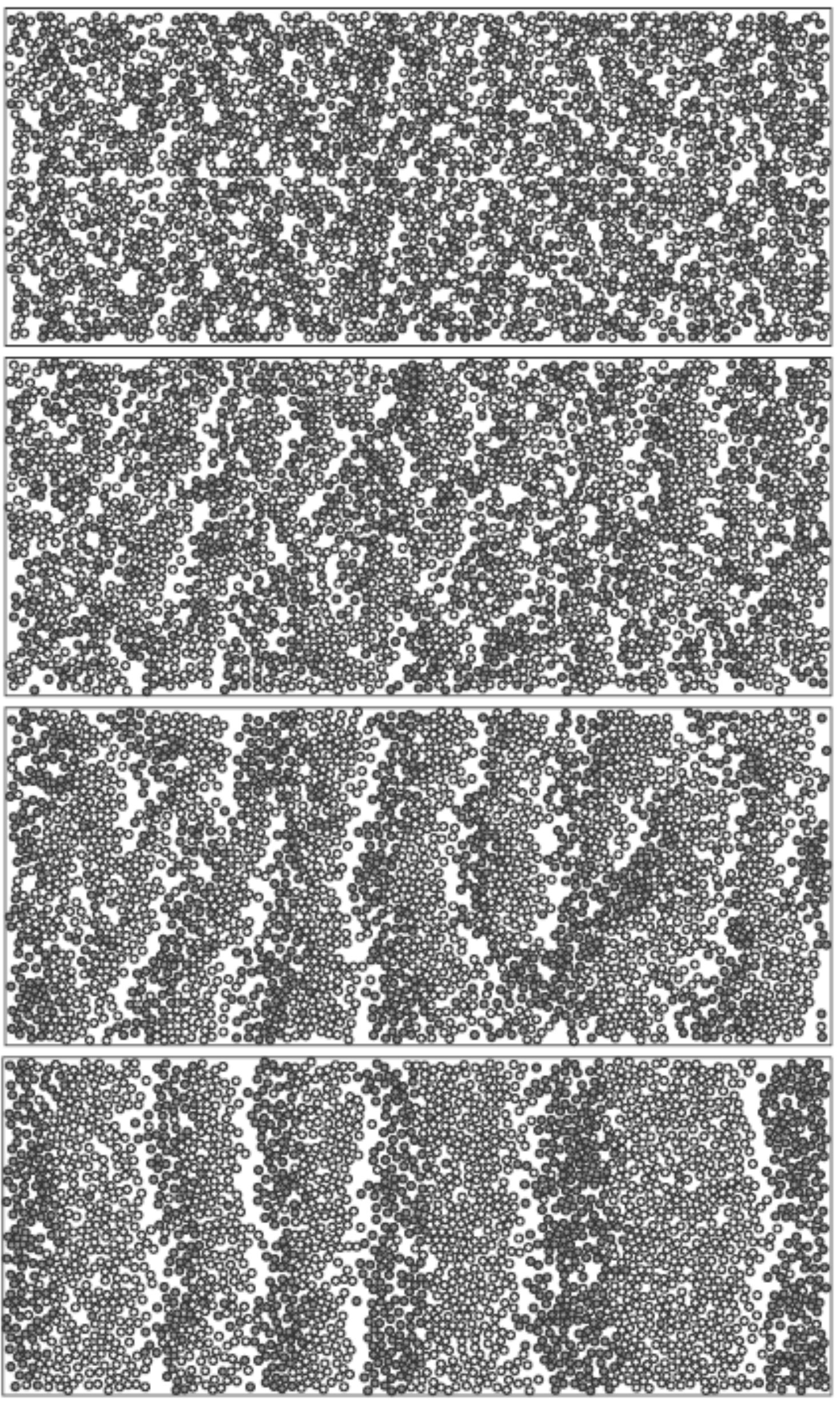}
\caption{Left: Pictures of the system studied in the experiment~\cite{M00}. Time runs from the top to the bottom and
the oscillation direction of the plate is perpendicular to the stripe orientation. Right: Molecular dynamics simulations of the same system~\cite{PCSNC07}
(stripes are perpendicular to the oscillation direction). Reprinted figure with permission from~\cite{M00}. Copyright (2000) by the
American Physical Society.}
\label{fig_stripes}
\end{figure}

We first consider a mixture of two kinds of particles: poppy seeds and
copper spheres~\cite{M00}. The initially homogeneous mixture is a
mono-layer placed on a plate which is horizontally oscillated, with
amplitude $A$ and frequency $\omega$. For packing fractions greater
than a threshold value, the two components are observed to segregate
in stripes of the two components, perpendicular to the oscillation
direction of the plate, see left panel of Figure~\ref{fig_stripes}.
The evolution in time of these patterns presents a coarsening
dynamics: the total number of stripes decreases while their average
amplitude, $L(t)$, increases  with time $t$. The behavior turns
out to be well described by a power law, $L(t)\sim t^z$, with
dynamical exponent $z=1/4$. A stochastic model based on a random walk,
proposed in the geological context of stone striping~\cite{M94}, can
explain the observed phenomenon.  Let us consider the stripe
  distribution $f(x,t)$, that represents the number of stripes of
  width $x$ at time $t$.  The evolution of the stripe width occurs by
  particles diffusing between stripes and it is described by a
  diffusion equation: $\partial f(x,t)/\partial t\propto
  \nu(t)\partial^2 f(x,t)/\partial x^2$. Here the frequency $\nu(t)$
  varies in time beacause particles have to diffuse in the region
  between the stripes to produce a fluctuation. This region is
  expected to scale as the average width of stripes $L(t)\equiv
  \langle x(t)\rangle$, and therefore $\nu(t)\propto L(t)^{-2}$.
  Introducing the variable $\tau=\int \nu(t)dt$, the diffusion
  equation is rewritten as $\partial f(x,\tau)/\partial
  \tau\propto \partial^2 f(x,\tau)/\partial x^2$, which gives $\langle
  x(\tau)^2\rangle\sim \tau$, and thus, using the defintion of the
  variable $\tau$, one readily gets the behavior $L(t)\propto
  t^{1/4}$.  In a successive experiment~\cite{RM02}, the behavior of
the same system has been investigated as a function of the packing
fraction, leading to the identification of a continuous phase
transition for the segregation phenomenon.

\begin{figure}
\includegraphics[width=7cm,clip=true]{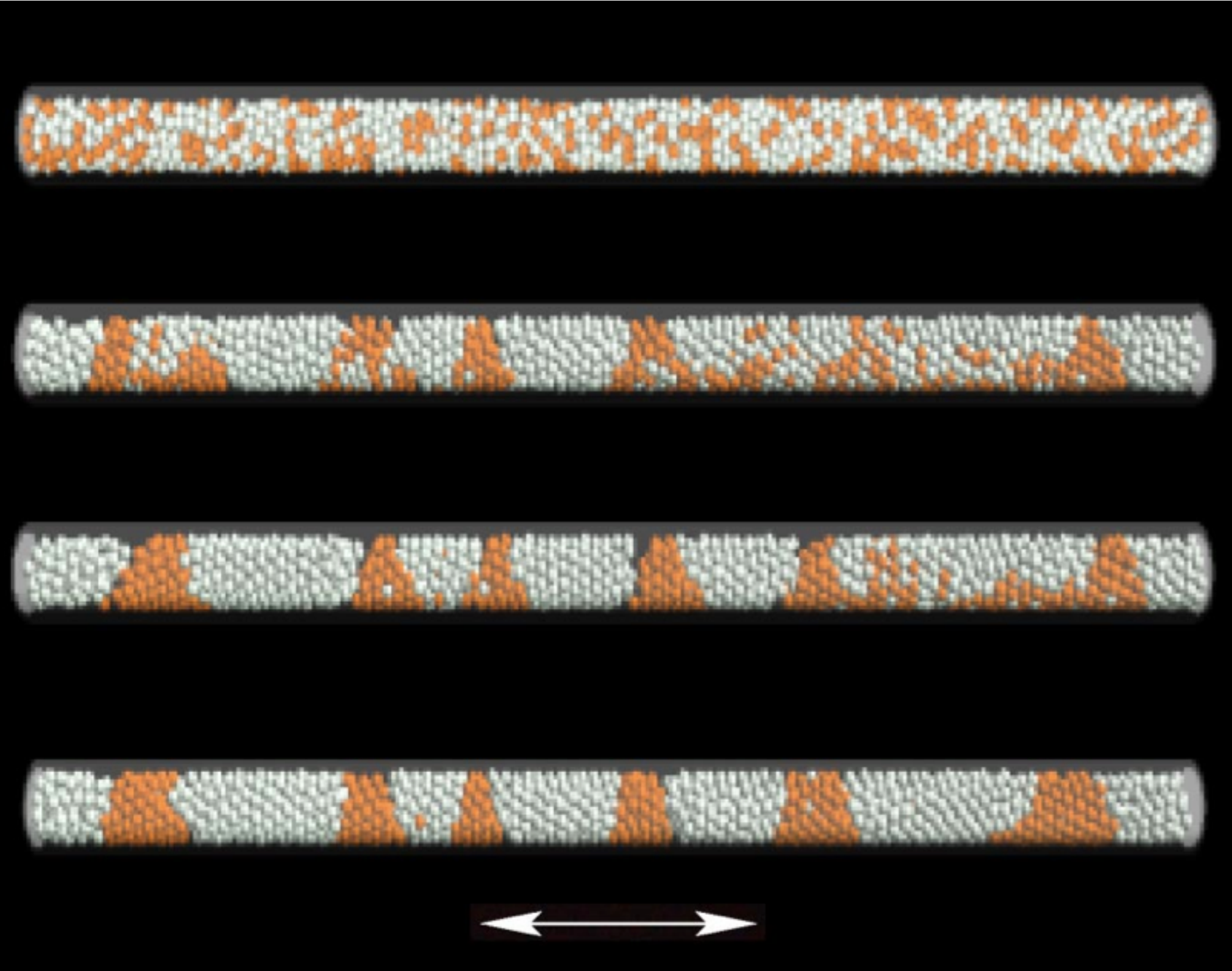}
\caption{Numerical simulations reproducing experiments on the
  segregation dynamics of a granular mixture immersed in a fluid,
  from~\cite{SSK04}.  Time runs from the top to the bottom and stripes
  are perpendicular to the oscillation direction of the cylinder 
    (indicated by the double arrow). Reprinted figure with permission
  from~\cite{SSK04}. Copyright (2004) by the American Physical Society.}
\label{fig_stripes2}
\end{figure}

Segregation via stripe coarsening is also observed in granular
mixtures immersed in fluids, subjected to horizontal vibration, as
reported in~\cite{SSK04}, see Figure~\ref{fig_stripes2}. The two
components in this case differ for masses and viscous coefficients,
and the force exercised by the fluid produces a differential
drag, which is responsible for the segregation mechanism. 

Numerical studies via molecular dynamics
simulations~\cite{PCCN05,PCSNC07} of a two-dimensional mixture of
granular disks, inspired by the first experiment described above, support
the interpretation that the different drag forces acting on the
components of the mixture induce the segregation process.  Such
simulations reproduce the same phenomenology observed in the
experiments, see right panel of Figure~\ref{fig_stripes}. Since the
mixture components have different friction coefficients and masses,
their dynamics is ruled by different relaxation times. Therefore,
different particles tend to oscillate with different amplitudes and
phases, producing the segregation in a stripe pattern.

An accurate theoretical description of these systems can be attempted by means of
an effective interaction theory~\cite{PCCN06}.  This approach allows
one to study an out-of equilibrium dissipative driven system in terms
of an ``equilibrium'' mono-disperse system. For a granular mixture
subject to horizontal oscillations, it has been shown that the
effective interaction force is anisotropic and presents a repulsive
shoulder at long distances which is prominent in the direction of
oscillation. This description leads to a phenomenological
Cahn-Hilliard equation which reproduces the observed phenomenology and
clarifies the origin of the differential drag mechanism for
spontaneous segregation and coarsening in these systems.

\paragraph{Rotating tumblers.} A different experimental setup where the dynamics of granular mixtures
is often studied is represented by rotating tumblers (of circular or
square section). Several possible segregation patterns are observed,
e.g. axial banding, radial streaks, etc., all displaying slow domain
growth. For a recent detailed review on this specific topic, we refer
the reader to~\cite{ST11}. The mixture segregation appears to occur in three stages:
first radial segregation is observed, then axial segregation takes place, with the formation
of bands which eventually coarsen slowly with the number of tumbler rotations.  
Here we focus on three experimental studies
in order to illustrate this general phenomenology: the first two are
concerned with the phenomenon of axial segregation in long tumblers, while the last one
with the observation of coarsening of radial streaks in a quasi-two-dimensional setup.

The first experiment analyzes the axial segregation of dry and wet
granular media, in circular or square tumblers~\cite{FO03}. If a
cylinder filled with a binary granular mixture is rotated around its
longitudinal axis, for specific rotational speeds, which play
  here the role of control parameter, one observes the formation of
alternating axial bands of the two components: initially, the
particles separate radially in the plane perpendicular to the rotation
axis, and later, a further segregation process occurs, where
coarsening of the bands takes place. More precisely, the number of
bands, $N$, grows in time at short times and then slowly decreases. At
long times, one observes for $N$ a logarithmic decay, as $-k \log(n)$,
where $n$ is the number of rotations. This kind of behavior can be
explained by some models for standard coarsening in one
dimension~\cite{B94} and it is also predicted by a continuous theory
for granular segregation~\cite{AT99}.

A detailed analysis of the specific mechanisms driving the coarsening
process in similar geometries for wet granular materials is reported
in~\cite{FVSNNS06}. In this experimental study, the authors
investigate, via optical methods and nuclear magnetic resonance
imaging, the dynamics of a binary mixture of glass beads of different
radii immersed in water, in a long cylindrical rotating drum. The
axial segregation phenomenon is found to exhibit a slow dynamics,
where the number of stripes decays logarithmically with the number of
rotations. A significant difference with respect to the dry case is
that the coarsening phenomenon is observed at lower rotation speeds in
the presence of water. The physical mechanism responsible for
coarsening is related to the redistribution of the small particles. However, a
clear understanding of the late stage of the coarsening dynamics in these systems is still lacking,
and the problem is the focus of very recent studies~\cite{FSS14}.

\begin{figure}
\includegraphics[width=11cm,clip=true]{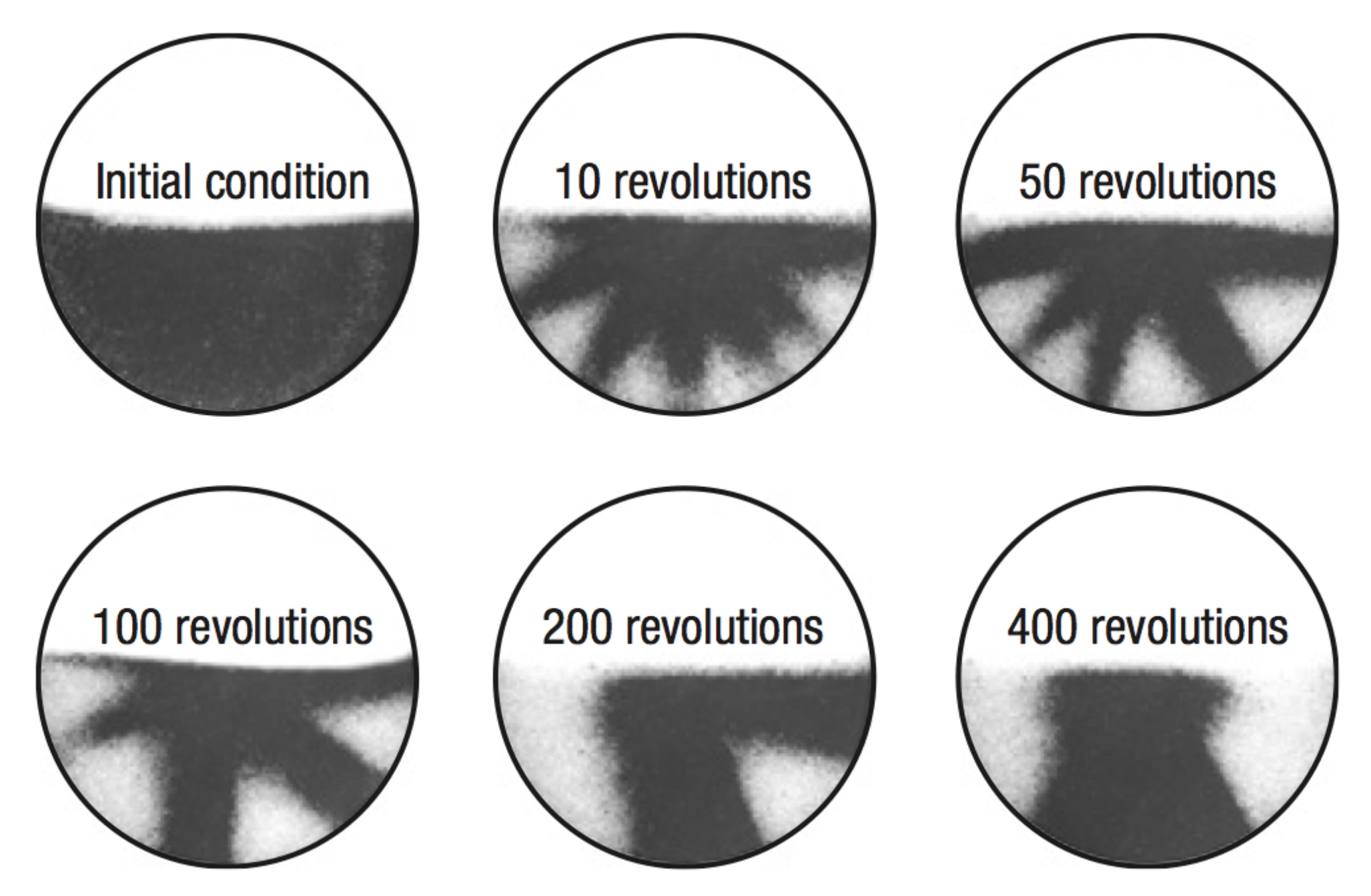}
\caption{Pictures of the experiment~\cite{MMBOL08}, where a quasi-2D
  circular tumbler, filled up to 55\% with a binary mixture of
  granular particles, is slowly rotated (2 revolutions per
  minute). The radial streaks are observed to coarsen into one,
  producing a final segregation of the components (notice that in
    an angular phase space coarsening cannot last forever and a
    complete segregation is expected). Reprinted by permission from Macmillan Publishers Ltd: Nature Physics~\cite{MMBOL08}, copyright (2008).}
\label{fig_stripes3}
\end{figure}

A different geometry used to study mixture segregation consists of a
quasi-two-dimensional slowly rotating tumbler, usually filled up to
$\sim 50\%$ with particles. In the experiments reported
in~\cite{MMBOL08}, the very long time coarsening regime, leading to a
complete final segregation, was studied (see Figure~\ref{fig_stripes3}). The experiment investigates the
dynamics of a binary granular mixtures of glass particles, differing
in size or in size and density. Even in this geometry, starting from a homogeneous state,
when a slow rotation is applied to the tumbler, the mixture initially
segregates into radial streaks, which eventually coarsen into one,
provoking a final separation of the components.

\section{Granular compaction}
\label{sec:comp}

\begin{figure}
\includegraphics[width=11cm,clip=true]{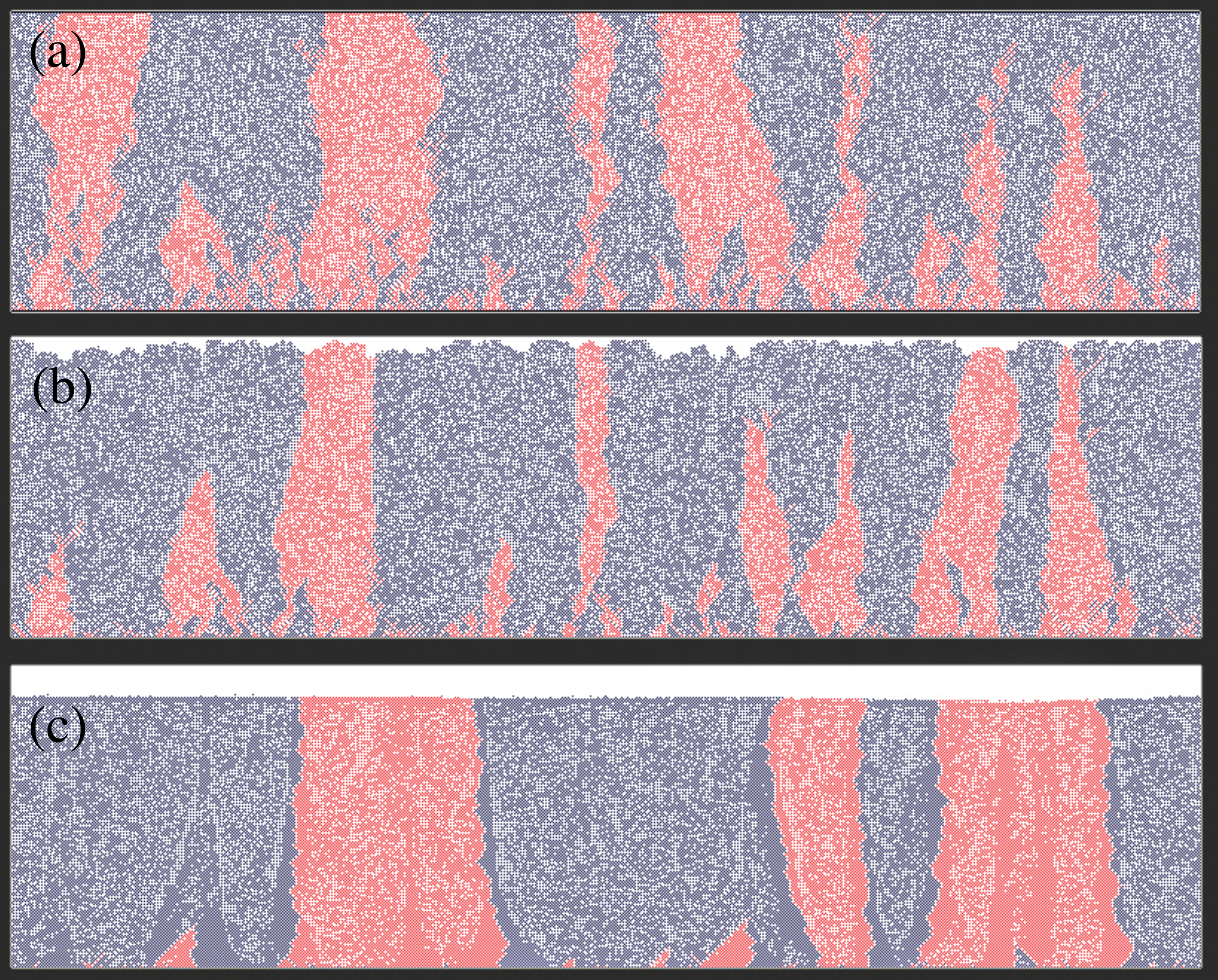}
\caption{Pictures of simulations from Tetris-like model for granular compaction~\cite{baldcomp}. The three panels reproduce the compaction dynamics at three different times $t_a<t_b<t_c$. Colored pixels are granular particles, while white pixels are voids. Note how the upper free surface of the granular media decreases in height, because of compaction. The colors of the pixels (red and blue) indicates the two different possible arrangements of the grains. An evident coarsening of domains appears. The average width of the domains grows at first as $t^{1/4}$, until all the domains span vertically the whole system. Then, a growth like $\sqrt{t}$ follows, due to the diffusion and annihilation of domain interfaces.}
\label{fig_tetris}
\end{figure}

Finally, in the pure solid phase, an interesting ordering phenomenon
is compaction~\cite{bencompx,bencomp}. This industrial relevant procedure
consists in a weak, sometimes sporadic, energy injection aimed at
increasing the packing fraction of the system. There is no general
optimal protocol to compact a granular medium, and the most accepted
theoretical scenario considers the dynamics of bubbles of mis-aligned
grains which evaporate, allowing the coalescence of optimally arranged
islands and a progressive reduction of total occupied
volume. In this context, many lattice models have been
proposed~\cite{baldcomp}, where the slow coarsening of domains of higher compactivity
is observed (see Fig.\ref{fig_tetris}). In order to discriminate different solid, mechanically
stable states of a granular, a general analogy with equilibrium
statistical physics has been proposed by Edwards~\cite{EO89}, where
energy is replaced by occupied volume, and the entropy is a measure of
the number of configurations in the position of grains which result in
the same occupied volume.

After an external perturbation, due to shaking or vibration, the
system organizes itself in a new metastable configuration. In this
way, it explores a phase space where each accessible point corresponds
to a configuration mechanically stable. This observation suggested the
idea to introduce a formalism analogous to the standard statistical
mechanics~\cite{EO89}. The central hypothesis is that the system can
be described by an ensemble average over the mechanically stable
states (called ``inherent states''). Then, assuming that each state
has an equal probability a priori to occur, the probability $P_r$ to
find the system in a inherent state $r$ is obtained maximizing the
entropy $S=-\sum_r P_r\log P_r$, with the macroscopic constraint that
the total energy of the system $E=\sum_r P_r E_r$ is fixed. This gives
the Gibbs result $P_r\propto \exp(-\beta_{conf}E_r)$, where
$\beta_{conf}$ is a Lagrange multiplier called inverse configurational
temperature. A generalized partition function can be then
introduced. For example, in the explicit case of a monodosperse system
of hard spheres with mass $m$ in the gravitational field $g$, where
the centers of mass of grains are constrained to move on the sites $i$
of a cubic lattice, one has the Hamiltonian:
  $\mathcal{H}=\mathcal{H}_{hc}+mg\sum_iz_in_i$, where
  $\mathcal{H}_{hc}$ represents the hard-core interaction between
  grains, preventing the overlap of nearest neighbor grains, $n_i$ is
  the occupation variable on site $i$ and $z_i$ its
  height~\cite{FNC02}.  In order to validate the Edwards hypothesis
one has i) to introduce a dynamics allowing the system to explore the
phase space; ii) to verify that, at stationarity, the system
properties do not depend on the specific parameters of the dynamics;
and iii) to check that the temporal averages obtained from such a
dynamics are consistent with those obtained from the Gibbs
distribution. This approach turned out to be effective in several
idealized models for granular compaction~\cite{BFS02,CBL02,FNC02}.

\section{Conclusions}

In this short review, we have described some relevant examples where
coarsening phenomena are observed in the context of granular media.
Remarkably, the study of these non-equilibrium athermal systems
reveals that, in many cases, a dynamics analogous to the phase
ordering kinetics of standard critical phenomena takes place. Concepts
such as ``order parameter'' and ``phase transition'' turn out to be
effective to describe instabilities, clustering, mixture segregation,
compaction, and other observed behaviors.  This suggests that the
tools of statistical mechanics can be extended and generalized to a
wider context, where fluctuations are induced by external forcing,
rather than by thermal agitation, and dissipation drives the systems
out of equilibrium.  In that respect, the physics of granular matter
shares many features with other non-equilibrium systems, such as
glasses, suspensions, foams, traffic flow, active matter. In
  particular, in the context of self-propelled particles, the
  phenomenon of motility induced phase separation can be explained by
  the same models of coarsening used for describing vibrated granular
  media, see for instance~\cite{W14}.  These considerations
strengthen the idea that a general theory can be able to describe a
wide gamut of different physical systems, at least in those cases
where the emergent behavior can be characterized in terms of few
macroscopic parameters. Here, as in many aforementioned systems, a
derivation of the coarse grained models from the microscopic dynamics
is, in our opinion, an important step, deserving further
investigation.

\end{document}